\documentclass[12pt,a4paper]{article}
\usepackage{a4wide}
\usepackage{amsfonts}
\usepackage{amssymb}
\begin{document}

\def\be{\begin{equation}}
\def\ee{\end{equation}}
\def\re{(\ref }
\def\tr{{\rm tr \, }}

\let\a=\alpha \let\b=\beta \let\g=\gamma \let\d=\delta
\let\e=\varepsilon \let\ep=\epsilon \let\z=\zeta \let\h=\eta \let\th=\theta
\let\dh=\vartheta \let\k=\kappa \let\l=\lambda \let\m=\mu
\let\n=\nu \let\x=\xi \let\p=\pi \let\r=\rho \let\s=\sigma
\let\t=\tau \let\o=\omega \let\c=\chi \let\ps=\psi
\let\ph=\varphi \let\Ph=\phi \let\PH=\Phi \let\Ps=\Psi
\let\O=\Omega \let\S=\Sigma \let\P=\Pi
\let\Th=\Theta \let\L=\Lambda \let \G=\Gamma \let\D=\Delta

\def\({\left(} \def\){\right)} \def\<{\langle} \def\>{\rangle}
\def\lb{\left\{} \def\rb{\right\}}
\def\[{\lbrack} \def\]{\rbrack}
\let\lra=\leftrightarrow \let\LRA=\Leftrightarrow
\def\ul{\underline}
\def\wt{\widetilde}
\let\Ra=\Rightarrow \let\ra=\rightarrow
\let\la=\leftarrow \let\La=\Leftarrow

\def\CG{{\cal G}}\def\CN{{\cal N}}\def\CC{{\cal C}}
\def\CL{{\cal L}} \def\CX{{\cal X}} \def\CA{{\cal A}} \def\CE{{\cal E}}
\def\CF{{\cal F}} \def\CD{{\cal D}} \def\rd{\rm d}
\def\rD{\rm D} \def\CH{{\cal H}} \def\CT{{\cal T}} \def\CM{{\cal M}}
\def\CI{{\cal I}}
\def\CP{{\cal P}} \def\CS{{\cal S}} \def\C{{\cal C}}
\def\CR{{\cal R}}
\def\CO{{\cal O}}

\newcommand*{\dId}{{\mathchoice%
{\mathrm{1\mkern-4.3mu I}}%
  {\mathrm{1\mkern-4.3mu I}}%
{\mathrm{1\mkern-3.1mu I}}%
  {\mathrm{1\mkern-3.1mu I}}}}
\newcommand*{\dR}{{\mathbb R}}
\newcommand*{\dN}{{\mathbb N}}
\newcommand*{\dZ}{{\mathbb Z}}
\newcommand*{\dC}{{\mathbb C}}
\newcommand{\Diff}{\mbox{\rm Diff}}
\newcommand{\diff}{\mbox{\rm diff}}
\newcommand{\semidir}{\mathrm{%

\times\mkern-3.3mu\protect\rule[0.04ex]{0.04em}{1.05ex}\mkern3.3mu\mbox{}}}
\newcommand{\dd}[1]{\frac{d}{d#1}}
\newcommand{\pd}[1]{\frac{\partial}{\partial #1}}
\newcommand{\zq}{\overline{z}}
\def\tl{{\widetilde \lambda}}

\newtheorem{theo}{Theorem}
\newtheorem{lemma}[theo]{Lemma}

\def\2{\frac12}
\def\pl{\plabel}
\def\Pf{{\it Pf\/}}
\def\Ad{{\it Ad\/}}
\def\Lieg{{\it Lie\/}G}
\def\tr{{\it tr\/}} 
\def\ad{{\it ad\/}}
\def\ba{\begin{eqnarray}}
\def\ea{\end{eqnarray}}
\def\bbbc{{\mathchoice {\setbox0=\hbox{$\displaystyle\rm C$}\hbox{\hbox
to0pt{\kern0.4\wd0\vrule height0.9\ht0\hss}\box0}}

{\setbox0=\hbox{$\textstyle\rm C$}\hbox{\hbox
to0pt{\kern0.4\wd0\vrule height0.9\ht0\hss}\box0}}
{\setbox0=\hbox{$\scriptstyle\rm C$}\hbox{\hbox
to0pt{\kern0.4\wd0\vrule height0.9\ht0\hss}\box0}}
{\setbox0=\hbox{$\scriptscriptstyle\rm C$}\hbox{\hbox
to0pt{\kern0.4\wd0\vrule height0.9\ht0\hss}\box0}}}}
\def\empty{{\emptyset}}

\newcommand{\proofend}{\raisebox{1.3mm}{%
\fbox{\begin{minipage}[b][0cm][b]{0cm}\end{minipage}}}}
\newenvironment{proof}[1][\hspace{-1mm}]{{\noindent\it Proof #1:}
}{\mbox{}\hfill \proofend\\\mbox{}}

\noindent
FSUJ-TPI-09/99     
{\renewcommand{\thefootnote}{\fnsymbol{footnote}}
\hfill   DESY 00-012   \\[2mm]
ESI 831 (2000) \hfill \phantom{hep-th/0001141}\\
\medskip
\begin{center}
{\LARGE \bf
Closed constraint algebras and \\[4mm] path integrals for 
loop group actions}  

\vspace{1.5em}
Anton Alekseev\footnote{e-mail: {\tt alekseev@teorfys.uu.se}},\\
Institutionen f\"or Teoretisk Fysik, Uppsala Universitet, \\
Box 803, S--751 08 Uppsala, Sweden\\[3mm]
Volker Schomerus\footnote{e-mail: {\tt vschomer@x4u.desy.de}},\\
II. Institut f\"ur Theoretische Physik, Universit\"at Hamburg,\\
Luruper Chaussee 149, D--22761 Hamburg, Germany\\[3mm]
Thomas Strobl\footnote{e-mail: 
{\tt Thomas.Strobl@tpi.uni-jena.de}},\\
Institut f\"ur Theoretische Physik, Universit\"at Jena\\
D--07743 Jena, Germany\\
\vspace{1.5em}
\begin{abstract}
In this note we study systems with a closed algebra of second class 
constraints. We describe a construction of the reduced theory that 
resembles the conventional treatment of first class constraints. It 
suggests, in particular, to compute the symplectic form on the reduced 
space by a fiber integral of the symplectic form on the original space.
This approach is then applied to a class of systems with loop group
symmetry. The chiral anomaly of the loop group action spoils the 
first class character of the constraints but not their closure. 
Proceeding along the general lines described above, we obtain a 
2--form from a fiber (path)integral. This form is not closed as a 
relict of the anomaly. Examples of such reduced spaces are provided 
by D-branes on group manifolds with WZW action. 
\end{abstract}
\end{center}}
\setcounter{footnote}{0}

\section{Introduction}
According to Dirac's classification, constraints in Hamiltonian 
mechanics split into first--class and second--class. 
The theory of the first--class constraints is  well developed 
because it is a major tool in gauge theories. Second--class 
constraints naturally arise in gauge theories with anomalies: 
quantum corrections may cause first--class constraints of 
the classical system to become second--class \cite{Faddeev}.

We reconsider Dirac's approach to second--class constraints 
and give a new realization of the reduced phase space which is 
more in line with the  reduction procedure for the first--class 
constraints. In this framework the Liouville form on the reduced 
phase space can be obtained by fiber integration from the Liouville 
form on the original phase space of the system. \footnote{We use 
the term `Liouville form' for the exponential of a symplectic 
form}. The procedure can suffer from possible global difficulties 
similar to the Gribov problems one often encounters in the context 
of ordinary (anomaly--free) gauge theories.

Our main interest is to apply this formalism to loop group actions on
symplectic manifolds. If the Poisson bracket of the symmetry generators 
contains a Schwinger term (for instance, this is the case in the WZW 
model), the constraints become second--class.  Such a situation was 
considered in the mathematical literature \cite{AMM},\cite{AMW}. We 
use a fiber integration procedure to derive the Liouville form on the 
reduced phase space. As a new manifestation of the anomaly it turns 
out that this form is not closed (see also \cite{AMW})!

`Anomalous' reduced spaces of this type naturally arise in the theory 
of D--branes on group manifolds \cite{AS}. It is an interesting problem 
to develop a consistent quantization theory for such spaces. Because of 
the connection between deformation quantization and open strings 
(see e.g.\ \cite{Kon,CaFe,S}) one expects valuable insights from 
open string theory. For the case of D--branes on group manifolds this 
was analysed in \cite{ARS}.

\section{Dirac brackets from fiber integration}

The aim of the present section is to reformulate the symplectic 
reduction for a system of constraints which form a closed algebra. 
To begin with we shall briefly recall the standard theory of 
reduction. This is used as a starting point for presenting an 
alternative formulation, applicable to closed constraint algebras 
under certain additional conditions. For the case of a purely 
second--class constraint system the approach suggests to 
construct the symplectic two--form and its associated Liouville 
form through a fiber integral. The material within this 
section serves as a toy model for the discussion of infinite 
dimensional phase spaces with an anomalous loop group action
in Section 3 and 4.
\medskip

Let us denote the original, unconstrained phase space by 
$N$ and let $x^i$, $i = 1, \dots, 2n$, be local coordinates.  
Their Poisson bracket is denoted by $P^{ij}= \{ x^i , x^j \}$ 
and its associated symplectic form  by   
$$
\Omega \ =\  \frac{1}{2} \Omega_{ij} \ dx^i dx^j \ \ .
$$
$\Phi^{\alpha}= \Phi^{\alpha}(x)$,
$\alpha = 1, \dots, 2m$, are constraints in this phase space. 
According to our assumptions they form a closed algebra, i.e.\ 
\be  
\{ \Phi_\alpha , \Phi_\beta \} = \Pi_{\alpha \beta}(\Phi)
\label{closed} \, \, 
\ee
with a matrix $\Pi$ that is a function of the constraints 
$\Phi_\alpha$ only. For simplicity we also assume that the 
constraints are independent from one another ({\em irreducible} 
constraints) and that they are all {\em regular} so that they 
may be used as local coordinates in phase space, at least in 
a neighborhood of the constraint surface $\Phi(x)=0$. Details
and further results used is the text below can be found, e.g.\ , 
in \cite{Henneaux}.
\smallskip

The standard procedure of symplectic reduction proceeds as 
follows: First the symplectic form $\O$ is pulled back to 
the constraint surface, which we denote by $N_0$. The resulting 
two--form $\O_0$ on $N_0$ is degenerate in general. Its 
kernel is surface--forming, however, and the quotient of 
$N_0$ with respect to the orbits (``gauge orbits'') is the 
reduced phase space $M$. By construction, the induced two--form
$\o$ on $M$ is nondegenerate. 

In the case of mere second--class constraints, i.e.\ $\ \det \Pi 
\neq 0 $ \ on $N_0$ in our context, the last step does not arise, 
since $\O_0$ is nondegenerate already. The Poisson bracket 
associated with $\O_0$ may be obtained directly from the Poisson 
bracket on $N$ by the following prescription  
$$
\{ f, g \}_D \ :=\  \{ f, g\} - \{ f, \Phi^{\alpha} \} 
(\Pi^{-1})_{\alpha \beta}
\{ \Phi^\beta, g \} \ \ , 
$$
which is defined at least in some neighborhood of the constraint 
surface $N_0 \subset N$. As a bivector--field this bracket, known 
as the Dirac bracket,  is {\em tangential\/} to the constraint 
surface $\Phi(x)=0$ (in contrast to the original Poisson bracket). 
Hence, it has a push--forward to $N_0$, and this coincides with 
the inverse of $\O_0$. 

Due to the closedness of our constraint algebra \re{closed}), the 
Hamiltonian vector fields 
$$ v^\a :=  \{ \Phi^\a, \cdot \} \equiv P^{ij} \ 
\frac{\partial \Phi^\alpha}{\partial x_i} \
 \frac{\partial}{\partial x_j} $$
are surface--forming everywhere in $N$.\footnote{For this the 
assumption on the closure of the constraint algebra is necessary 
as one can see e.g.\ from the fact that even a set of first--class 
constraints is surface--forming only along the constraint surface, 
in general.} In fact, one can easily show that  
$$
[v^\alpha, v^\beta]\ =\  
\frac{\partial \Pi^{\alpha \beta}}{\partial \Phi^\gamma} \ 
v^\gamma
\ \ .
$$
Thus the $v^\a$s generate orbits in {\em all\/} of $N$ and this 
is true even if the set $(\Phi^\a)$ contains second--class 
constraints. 
\medskip

The last observation opens new possibilities for performing the 
symplectic reduction, viewing the reduced phase space $M$ as an 
appropriate orbit space. The difference to the standard Dirac 
procedure outlined above lies primarily in the treatment of the 
second--class constraints, which are dealt with very analogously 
to first--class constraints in their standard reduction. 
Consequently, our main focus in the remainder of this section 
will be on second--class constraints. We will briefly comment  
on the extension to more general cases with the simultaneous 
presence of first-- and second--class constraints towards the 
end of the section.  

For the case of pure second--class constraints, the matrix 
\re{closed}) is nondegenerate on the constraint surface, i.e.\ 
at $\Phi=0$. In what follows we will strengthen this requirement 
by assuming that $\det \Pi(\Phi) \neq 0$ not only at the value 
zero,  but for all values of $\Phi$ adopted. 
This permits us to regard the image $C$ of the constraint map 
$\phi: N \to C \subset \dR^{2m}$, $x \mapsto \Phi^\a(x)$, as a 
symplectic manifold; indeed $C \equiv$ Im$\phi$ is
endowed naturally with the symplectic form 
\be \varpi\ =\ \frac{1}{2}
(\Pi^{-1})_{\alpha \beta} \ d\Phi^{\alpha} d\Phi^{\beta} \ \ .
\label{varpifinite} \ee  
By construction, the map $\phi$ is Poisson. 

In contrast to the standard approaches in which the reduced 
phase space $M$ is regarded as a restriction of the original 
phase space $N$ to the constraint surface $\Phi(x)=0$, we 
propose to view $M$ as the {\em space of orbits\/} generated 
by the second--class constraints $\Phi^\a$. Since the 
constraints are second--class, their Hamiltonian vector fields 
$v^\a$ are nowhere tangential to the constraint surface. Thus, at 
least locally, any point of the constraint surface, i.e.\ of the 
reduced phase space $M$, corresponds to an orbit (namely the one 
that is generated by the $v^\a$s through the point in question). 
\medskip

Before we follow this general idea, we shall pause for a moment 
and comment the possible global difficulties. The full equivalence 
between the reduced phase space $M$ and the orbit space requires 
any orbit to intersect the constraint surface once and only once. 
Despite the fact that we required $\Pi$ to be nondegenerate
everywhere, the orbits do not necessarily have this property in
general. The situation we meet here is similar to the one of 
choosing gauge conditions for a set of first--class constraints, 
with $\det \Pi$ playing the role of the Faddeev--Popov determinant. 
Note that the combined system of first--class constraints and 
gauge--conditions forms a set of second--class constraints.  It 
is known that even for a nonvanishing Faddeev--Popov determinant 
(along the intersection of the constraint surface with the gauge 
conditions) the chosen gauge conditions may show global deficiencies, 
in which case they are referred to as having a Gribov problem 
\cite{Gribov}. 

By analogy, we call the orbits generated by the $v^\a$s to have a
{\em Gribov problem}, if they do not intersect the constraint 
surface precisely once. To conclude these remarks, let us 
illustrate such problems through the following simple example 
where we take $N = T^*\dR \setminus {(0,0)}$ 
with standard symplectic form $\O=dq \wedge dp$. Now let us choose the
constraints $$ \Phi^1 \ :=\ \frac{(q^2-p^2)}{2\sqrt{q^2+p^2}}-\frac12
\ \ \ \mbox{ and } \ \ \  
    \Phi^2 \ := \ \frac{qp}{2 \sqrt{q^2+p^2}}\ \ . 
$$ 
Their Poisson bracket is given by $\Pi^{12} = 1$ and one can easily 
establish that there is just {\em one\/} orbit in $T^*\dR 
\setminus {(0,0)}$. On the other hand, $T^*\dR \setminus {(0,0)}$ contains 
{\em two\/} points of the reduced phase space: $(q,p)=(\pm 1,0)$. 
In fact, the map from $N= T^*\dR \setminus {(0,0)}$ to $C$ defines 
a two--fold covering (as one may see most easily in polar 
coordinates). These constructions are easily extended to obtain 
examples with an arbitrary number of Gribov copies.
\medskip

In the absence of a Gribov problem, however, the reduced phase 
space $M$ may be fully identified with the space of orbits. Let 
$\pi$ denote the projection from $N$ to $M$ along the orbits. 
We then have the following proposition: The symplectic 
two--form $\omega$ on $M$ satisfies
\be 
\pi^* \omega\ = \ \Omega - \phi^* \varpi \ \ . \label{split}
\ee
The proof proceeds in several steps. First, we show that the 
form $(\Omega - \phi^* \varpi)$ descends to the space of leaves 
of the foliation. Indeed, it is horizontal,
\be
\Omega(v^\a, \cdot) -
\phi^* \varpi(u^\a, \cdot) \ = \ d \Phi_\a(x) - \phi^* d\Phi_\a\ 
=\ 0 \, . \label{proof1} 
\ee
Here $u^\a$ denotes the projection of the vector fields $v^\a$ to 
$C$ (which is well--defined since the $v^\a$s are tangential 
to the orbits): $u^\a = \{ \Phi^\a , \cdot \}_C$, the index $C$ being
used to make clear that the bracket corresponds to
\re{varpifinite}). For later use we remark here that by assumption on
the determinant of $\Pi$, the vector fields $u^\a$ --- and thus also
the vector fields $v^\a$ --- are nonzero everywhere; correspondingly,
the action generated by the constraints is free. 

Equation \re{proof1}) implies that, since the form in question is
closed, it is also invariant with respect to the flows generated by
the constraints. Hence, it is a pullback of some two--form on $M$, which
we denote by $\omega$.

Next, we show that $\omega$ coincides with the inverse of
Dirac's bracket. For this purpose we consider two functions
$f$ and $g$ on $N$ which are constant on the leaves of
the foliation. This implies that their Poisson
brackets with the constraints vanish, yielding 
$\{ f,g \}_D \ = \ \{ f, g\}$.
Denote the corresponding Hamiltonian vector fields
by $v_f$ and $v_g$. Note that one can use either the original 
Poisson bracket or the Dirac bracket to define them.    
We would like to show that
$$
\omega(\pi_*v_f , \pi_*v_g)\ =\ \{ f,g\}_D \, .
$$

Indeed,
$$
\{ f,g \}_D\ = \ \{ f , g \} \ = \ \Omega(v_f,v_g)\ =\ 
\omega(\pi_*v_f , \pi_*v_g)\ \ .
$$
Here we have used that the vector fields $v_f$ and $v_g$ project 
to zero by $\phi$. Thus we have fully established our formula 
(\ref{split}) above. 
\medskip 

In the presence of a Gribov problem with nonvanishing number of 
Gribov copies \footnote{One may also encounter situations in 
which some orbits do not intersect the constraint surface $\Phi(x)=0$
at all; for an illustration just add the constant value 2 to the 
first constraint in the example below.} the map $\phi$ restricted 
to an orbit is not injective. For the following constructions we 
shall assume that the restriction of $\phi$ is a bijection. Note 
that this property is not guaranteed by the absence of a Gribov 

problem, since $\phi$ may still fail to be surjective after 
restriction to an orbit. As an example we take $N$ to be $T^*\dR^2$ 
with the standard symplectic form and choose the constraints $\Phi^1 
:= \exp(q^2) \, \left[\exp(q^1)-1\right]$, $\Phi^2 := \exp(-q^1-q^2) 
\, p_1$,  which again leads to $\Pi^{12}=1$. Now, $C \equiv \mbox{Im}
\phi = T^* \dR$, but an orbit characterized by a fixed value of $q^2$ 
maps only to the parts of $C$ with $\Phi^1>- \exp(q^2)$. 
\medskip

If the map $\phi$ restricted to any orbit is surjective and 
there is no Gribov problem, the original phase space is a fiber 
bundle with typical fiber $C$ and base manifold $M$. In this 
case there is an alternative way to express the relation between 
the form $\Omega$ on $N$ and $\omega$ on $M$:
Let us consider the Liouville forms of mixed degree 
$L:=\exp(\Omega)$ on $N$ and  $l:=\exp(\omega)$
on $M$. The top degree components of $L$ and $l$ are the Liouville
volume forms on $N$ and $M$, respectively. We define the 
normalized fiber integral (or push forward map) $\pi_*$
over the leaves of our foliation by the formula
\be  \label{push}
\pi_* \alpha \ :=\  \frac{1}{{\rm Vol}\ C} \int_{\mbox{\it fiber}} \alpha \ \ .
\ee
Here $\alpha$ is a differential form on $M$ and
${\rm Vol}\  C$ is a (possibly infinite) symplectic volume
of the constraint space $C$. If $C$ is compact, $\pi_*$ is just
the ordinary push--forward map. Otherwise, the normalization 
factor $({\rm Vol}\  C)^{-1}$ is reminiscent of the infinite 
normalization constants in the definitions of path integrals.
By applying the fiber integral (\ref{push}) to the Liouville
form $L$, we obtain
\begin{eqnarray}
\pi_* L & = & \frac{1}{{\rm Vol}\ C} \int_{\mbox{\it fiber}} 
\exp(\Omega)=
\frac{1}{{\rm Vol}\ C} \exp(\omega) \int_{\mbox{\it fiber}} 
\exp(\Phi^* \varpi) 
  \nonumber\\[2mm]
& =&  \frac{\int_{C} \exp( \varpi)}{{\rm Vol}\ C} 
\exp(\omega)=l \ \ .
\label{fiber}
\end{eqnarray}

In the next section we shall generalize equation (\ref{split}) to an
infinite dimensional context where $\Omega$ is given by a symplectic
form on some space of fields. There will be one major difference in
comparison to the considerations in the present section: the forms
$\omega$ and $\varpi$ will no longer be closed! 
The fiber integral (\ref{push}) (which becomes a path integral) will
then provide a prescription of how to define the Liouville form $l$
for a nonclosed form $\omega$.
\medskip

Before turning to this, however, we briefly extend the 
above considerations to the general setting of a closed algebra 
of constraints, where there are both first-- and second--class 
constraints. Note in this context that although one may always 
replace a set of constraints by an equivalent set of constraints 
where first-- and second--class constraints are split 
\cite{Henneaux}, this splitting is achieved only on-shell 
(i.e.\ in a ``weak sense''). On the full phase of the original 
theory, however, it may be impossible to find a splitting 
for which the second--class constraints do not generate 
first-class constraints upon Poisson commutation. 

In the case of a closed constraint algebra containing first--class 
constraints, the matrix $\Pi(\Phi)$ is degenerate. Consequently, 
the manifold $C = \mbox{Im} \phi$ is no longer symplectic but only 
a Poisson manifold. Hence, $C$ foliates into symplectic leaves. 
Let $C_0$ denote the symplectic leaf containing the origin 
$\Phi =0$ and $\widetilde N_0$ be the pre-image of $C_0$, i.e.\ 
$\widetilde N_0 = \phi^{-1} (C_0)$. ($\widetilde N_0$ may be 
obtained equivalently through the action on $N_0$ of the flow 
generated by the constraints.)  The reduced phase space may now 
be regarded as the space of orbits in $\widetilde N_0$, at least 
in the absence of a Gribov problem. A formula of the type 
\re{split}) is true, if in the right--hand side $\O$ is the 
restriction of the symplectic form on the original space $N$ to 
$\widetilde N_0$ and $\varpi$ is the symplectic form on $C_0$. 
By means of such a formula one may, however, no more relate 
the Liouville forms on $N$ and $M$ such as in  \re{fiber}). The 
reason is that the fiber integration over $\O$ restricted to 
$\widetilde N_0$ yields zero, since in the presence of 
first--class constraints this differential form has a kernel 
along the fibers.

\section{Hamiltonian systems with loop group symmetry}

Now we turn to the infinite dimensional situation of interest.  Our
phase space $N$ is a field space with symplectic form $\Omega$. By
assumption, it has a Hamiltonian action of the loop group $LG$ of some
Lie group $G$, which we take to be compact, simple, and simply
connected for simplicity.  To an algebra element $\varepsilon(s) \in
L\CG$ we associate a Hamiltonian vector field 
\be v_\varepsilon = \{J_\varepsilon, \cdot\} \label{vectorfield}
\ee on $N$, where 
\be  J_\varepsilon\ =\ {\rm tr} \int_0^1 \varepsilon(s) J(s) ds 
\label{dual}  \ee
and $J(s)$ is a field giving rise to the moment map for the loop 
group action.

Using an orthonormal basis $t^a$ in the Lie algebra $\CG$, we can
write the Poisson brackets of the components of $J(s)$ in the form 
\be  \{ J^a(s), J^b(s') \} \ =\ k \, \delta^{ab}\, \delta'(s-s') + 
f^{ab}_c \,\delta(s-s')\, J^c(s) \ \ .  \label{currents} \ee 
Here $k$ is a coefficient in front of the anomalous term in the
bracket. We would like to use the currents $J^a(s)$ as constraints in
our Hamiltonian system.  If $k$ vanishes, they are first--class
constraints and can be treated by the standard procedure. Our main
interest is to deal with the case of nonvanishing $k$.  To simplify
notations we will set $k=1$ for the rest of the paper.

According to equation \re{currents}), the currents $J(s)$ form a
closed algebra of both first-- and second--class constraints. The zero
modes of the currents $J^a_0:= \int J^a(s) ds$ are first--class. All the 
remaining modes in a Fourier decomposition of $J(s)$ are second--class. 
The latter do not close among themselves since Poisson brackets of $J_n$ 
with $J_{-n}$ have $J_0$--contributions. Hence, the Fourier modes $J_n$ 
do not allow to split off a closed algebra of pure second class 
constraints. As we remarked above, such a splitting into closed 
first--class and closed second--class constraints need not even exist. 

In the present case, however, we can split the constraints into 
first-- and second--class. To see this, we return to the loop group 
$LG$, whose Lie algebra elements enter the
Hamiltonians \re{dual}).  $LG$ may be written as a semidirect product
of the group of based loops $\Omega G$, formed by the loops with
property $g(0)=e$, and the group $G$: Any $g(s) \in LG$ can be written
uniquely as $g(s) = {\widetilde g}(s) \, \hat g$ with ${\widetilde g}(s) \in
\Omega G$ and $\hat g \in G$. On the Lie algebra level this corresponds 
to the unique splitting of any $\e(s) \in L\CG$ into the sum of a
constant Lie algebra element $\e(0)$ and an ${\widetilde \e}(s) \in \O
\CG$: $\e(s)=\e(0) + {\widetilde \e}(s)$ with ${\widetilde
\e}(0)=0$. Re-expressing the relations \re{currents}) in terms of the
Hamiltonians \re{dual}), one finds
\be \{ J_\e, J_\eta \} = tr 
\int_0^1 \e(s) \eta'(s) ds + J_{[\e,\eta]} 
\label{currents2} \, \, . \ee
Since $\delta'(s-s')$ is an invertible operator on test-functions
vanishing on the endpoints of the interval, these relations become
second--class upon restriction to $\Omega \CG$. Moreover, the algebra of 
this subclass of Hamiltonians is obviously closed now. 
\medskip

So, following the ideas of section 2, we should now be able to forget
the first--class constraints and just restrict our attention to the
subclass of second--class constraints so as to perform the pushforward
integral \re{fiber}) we are after. However, at this point we have to
fight with the infinite dimensionality of the space of constraints
and with the properties of an (appropriately defined) dual for the 
Lie algebra of the group $\Omega G$. (Recall that the moment(um) map 
yields elements in the {\em dual\/} space of the Lie algebra of the 
group action in question, cf e.g.\ \cite{Woodhouse} for details.) 

In this paper, we do not intend to go into the functional analytical 
details that would be necessary to fully and rigoroulsy {\em extend\/} 
the approach of the previous section to the present infinite dimensional 
case (although this might yield interesting insights). Instead we will 
make use of a (mathematically rigorous) formula which is of the {\em 
form\/} of equation \re{split}) with a (weakly) nondegenerate $\varpi$, 
which, however, is not closed and thus not symplectic.  

For this purpose we return to the action of the group $\Omega G$.  As
follows from equation \re{currents}), this group (or also $LG$) acts
on the space of currents by standard gauge transformation, 
$$ J^g(s) \ = \ g^{-1}J\, g + g^{-1} \partial_s g \ \ .  $$ 
This action has no fixed points (here the restriction to $\O G$ becomes 
relevant!), similar to the flows of the vector fields $u_\a$ on $C$ in 
Section 2. Hence, the action of $\Omega G$ on $N$ is also free.  Then, 
one can form the space of orbits, $M:= N/\Omega G$ which replaces the 
space of leaves of the foliation of Section 2. The projection from $N$ 
to $M$ is denoted by $\pi$.

Similar to equation (\ref{split}) we may decompose the symplectic form 
$\Omega$ on the original according to (cf.\  Theorem 8.3 in \cite{AMM})
\be
\Omega \ = \ \pi^* \omega + J^* \varpi \ \ , \label{amm}
\ee 
where $\omega$ is a two--form on $M$ and $\varpi$ lives on the space
of currents (for a more precise definition of this space cf also
\cite{AMM}).  The explicit formula for $\varpi$ looks as follows. 
Denote by $\Psi$ the solution of the equation 
$$
\partial_s \Psi \Psi^{-1}\ =\ J(s) \ \ 
$$ 
with the boundary condition $\Psi(0)=e$.  In other words, $\Psi(s)$
is a path ordered exponential of $J(s)$ and $\Psi(1)$ is the holonomy
map, which takes values in the group $G$.  Obviously, this map
descends to $M$ and we shall denote the induced map by $\psi: M
\rightarrow G$.

The form $\varpi$ is given by
\be  \label{varpi}
\varpi\ :=\  \frac{1}{2} {\rm tr} \int_0^1 \left( \Psi^{-1} d \Psi \
\partial_s (\Psi^{-1} d \Psi) \right) \ \ .
\ee
As remarked above, it is not closed, 
$$ d \varpi \ =\ \frac{1}{6} {\rm
tr} ( \Psi^{-1}(1) d \Psi(1) )^3 \ \ .  $$ 
As a consequence of formula \re{amm}), the form $\omega$ is also not 
closed, 
$$ d \omega \ =\ - \frac{1}{6} {\rm tr} ( \psi^{-1} d \psi )^3 \ \ .  $$
Note, however, that the right--hand side of the last two formulas is
proportional to the coefficient $k$ in \re{currents}), which we have set to 
one thereafter. Thus, these forms become closed in the absence of the 
anomalous term in the current algebra. This observation will become relevant 
when interpreting the final result of the calculation in section 4. 


Although $\varpi$ is not symplectic, it comes very close to an inverse of the 
Poisson brackets \re{currents2}) between the second--class constraints. 
By straightforward calculation one verifies the two relations 
\be \iota(v_{\widetilde \e}) \varpi \equiv \varpi(v_{\widetilde \e} , 
\cdot ) \ = \ - d J_{\widetilde \e} \, \, , \qquad  
\varpi(v_{\widetilde \e} , 
v_{\widetilde \eta} ) = \{ J_{\widetilde \e} , J_{\widetilde \eta} \} \, . 
\label{wunder} \ee
For these relations to hold it is essential that one restricts the Lie
algebra elements to $\O \CG$ (for the corrections appearing otherwise
cf Proposition 8.1 in \cite{AMM}). In the finite dimensional setting,
equations of the form \re{wunder}) for a complete set of Hamiltonian
vector fields  are already sufficient to ensure that $\varpi$ is the 
sought--for symplectic form; the closedness would then follow 
automatically by validity of the Jacobi identity for the Poisson 
bracket. 

In the present infinite dimensional setting, the form $\varpi$
yielding relations of the form \re{wunder}) is even not unique. Indeed, 
one can change the splitting \re{amm}) by an arbitrary
2-form $\beta$ on the group $G$,
$$
\widetilde{\omega}=\omega+ \psi^* \beta \, , \qquad 
\widetilde{\varpi} = \varpi - \Psi^* \beta \ \ ,
$$ without affecting the relations \re{wunder}) where $\varpi$ is
replaced by $\widetilde \varpi$. Note that because the 3-form
$tr(\psi^{-1} d\psi)^3$ belongs to a nontrivial cohomology class on
$G$, also the 2-form $\tilde{\omega}$ is not closed, $$
d\tilde{\omega} = d\omega + \psi^* d\beta \neq 0.  $$
 
\medskip

The phase space $N$ is symplectic and carries the Liouville form
$L=\exp(\Omega)$. The (formal) top degree part of $L$ gives the
measure of the Hamiltonian path integral. Inspired by equation
(\ref{fiber}), we would like to (formally) {\em define\/} 
the Liouville form $l$ on $M$ by the formula 
\be  l:= \pi_* L \ =\ \frac{1}{\mbox{Vol} \O G}
\int_{\O G} \exp \O \ \equiv \ 
\frac{\int_{\Omega G} \exp(\varpi)}{{\rm Vol}\ \Omega G}\
\exp(\omega) \ \ , \label{refertoitlater}
\ee where we made use of the definition \re{push}) as well as of the
relation \re{amm}).  In the next section we compute the path integral
$$ I(\psi)\ :=\ \frac{1}{{\rm Vol}\ \Omega G} \int_{\Omega G}
\exp(\varpi) \ \ , $$ where $\psi= \Psi(1)$ is an element of $G$. Note
that the resulting integral will be a differential form of mixed degree 
rather than a function on $G$ (or its pullback to $M$).

\section{Evaluation of the path integral}

We want to integrate  $\exp(\varpi)$ over the group of based loops 
$\O G$. Therefore we split the field $\Psi(s)$ in formula \re{varpi}) 
into a product of an element $h \in \Omega G$ and an extra factor 
$\exp(\alpha s)$, i.e.\
$$     \Psi(s) \ = \ h(s) \exp(\alpha s)   \ \ , $$
where $h(s) \in G$ is a periodic $G$--valued function 
and $\a \in \CG$ is sent to the group element $\Psi(1) = \exp(\a) 
\in G$ by the exponential mapping.

A short and elementary computation allows to reexpress 
the form $\varpi$ in terms of the variables $h(s)$ and
$\alpha$. The result is, 
$$  
\varpi \ = \ \frac12 {\rm tr} \int_0^1 ds \left( 
           (h^{-1} dh) D_\a (h^{-1} d h) + 2 h^{-1} 
            dh d\a - d(e^{\a s}) e^{-\a s} d\a \right)\ \ . 
$$  
$D_\a$ denotes the covariant derivative $D_\a = \partial_s - 
\ad_\a$ where $\ad_\a (.) = [\alpha, .]$. The last term in 
$\varpi$ can be evaluated with the help of the following 
formula  
\be    \theta(s) \ : = \ d(e^{\a s}) e^{-\a s} \ = \ 
      \frac{1}{\ad_\a}( 1- e^{s \ad_\a} ) d \a\ \  .
 \label{lemm} \ee
Here, $1/\ad_\a = (\ad_\a)^{-1}$ is the inverse of the 
adjoint action $\ad_\a$ with $\a$. Note that the function 
$\frac{1}{x} (1-e^{sx}) = \sum_{n\geq 1} s^n x^{n-1}/n!$ is 
regular even at $x=0$ so that the right hand side of formula 
\re{lemm}) is well--defined. 
To establish eq.\ \re{lemm}) we differentiate the function 
$\theta(s)$ with respect to $s$ to find, 
$$ \partial_s \theta(s) \ = \ d \a + [\a, \theta(s)]\ \ . $$
If the ansatz $\theta(s) = \exp(s \ad_\a) \vartheta(s)$ is 
inserted into the expression for $\partial_s \theta(s)$ we 
deduce
$$ \partial_s \vartheta(s) = e^{-s \ad_\a} d\a \ \ . $$
This equation can easily be integrated to give the claimed 
formula for $\theta(s)$.
\medskip

Formula \re{lemm}) actually allows to perform the integral 
over $s$ for the third term in $\varpi$. This results in 
\be \varpi \ = \ \frac12 {\rm tr} \int_0^1 ds 
   \left( \phi D_\a \phi + 2 \phi d\a \right)  
    - \frac12 {\rm tr}\left(d\a \frac{1}{(\ad_\a)^2}
     (e^{\ad_\a} - 1 - \ad_\a ) d\a \right) \ \ . \label{v2} \ee
Again, the argument of the second trace is well--defined 
on the kernel of  $ad_\alpha$. In this expression for the form 
$\varpi$ we also introduced the field $\phi(s) = h^{-1}(s) 
d h(s)$. By construction, $\phi(s)$ is a fermionic field 
subject to the constraint $\phi(0) = 0 = \phi(1)$. The 
integral over the exponential of the two--form $\varpi$ is 
now reinterpreted as a fermionic ``path integral'' $\int 
\CD \phi \exp(\varpi)$. 


{}From the proof of eq.\ \re{lemm}) above it is obvious that 
$D_\a \theta(s) = d \a$. Therefore one can rewrite the form 
$\varpi$ also as
$$ 
\varpi \ = \ \frac12 {\rm tr} \int_0^1 ds 
   \left( \phi + \th \right) D_\a  \left( \phi + \th \right) \, . 
$$
This may lead one to conclude that the integration of $\exp \varpi$ 
over $\phi$ merely results in the Pfaffian of $D_\a$. However, $(\phi +
\th)(s)$ does not vanish at $s=1$ and a change of variables to $\phi +
\th$ is illegitimate.

We therefore proceed with integrating $ \exp(\varpi)$ in the form 
of eq.\ \re{v2}). As the last term does not depend on $\phi$ (resp.\ $h$)
and as, being a two--form, it commutes with the first two terms, we can 
split the exponential into two parts, the second one of which we may  
pull out of the integral, i.e.\  we shall write $\varpi = \varpi_1
- \varpi_2$ with 
\ba 
\varpi_1 & = & \frac12 {\rm tr} \int_0^1 ds 
   \left( \phi D_\a \phi + 2 \phi d\a \right) \ \ , 
      \ \ 
\label{vp1}
\\[2mm]
   \varpi_2 & = & \frac12 {\rm tr}\left(d\a \frac{1}{(\ad_\a)^2}
     (e^{\ad_\a} -1- \ad_\a ) d\a \right) \ \ 
\label{vp2}  
\ea
and compute the Integral $I = \int \CD \phi \exp(\varpi_1)$,  
leaving out the extra factor $\exp(-\varpi_2)$ for the moment. 
\medskip

The field $\phi(s)$ is a periodic fermionic field which admits the Fourier 
decomposition: $\phi(s) = \sum_n \phi_n \exp(2 \pi i n \, s)$. 
In terms of the Fourier modes $\phi_n$, the constraint $\phi(0) = 0$ 
becomes $\sum_n \phi_n = 0$. To turn the integral into a Gaussian 
one over {\em unrestricted}\/ variables, we introduce a Lagrange 
multiplier $\l$.  This leaves us with the computation of the following 
integral: 
$$ 
I \ = \ \int \prod_n d \phi_n d\l \,\,  
\exp{ \, {\displaystyle{\rm tr} \left( \2 \sum_m \phi_{-m} D_m
  \phi_m +  \phi_0 d \a + \lambda \sum_m \phi_m \right)}}  
\label{I1} 
\,  \, , 
$$
where $D_n \equiv 2 \pi i \, n - \ad_\a$ and $\l$ is a fermionic 
variable too. A product over the Lie algebra indices of $\phi_n$ and 
$\l$ in the integration measure is understood, furthermore. 
Defining $J_n = \l + \d_{n,0} \, d \a$, the second and third 

term in the exponent may be combined into $\sum_m \phi_{-m} J_m$.  

We would like to remark that the last reformulation of our 
integral involves the choice of some particular (anti)self--adjoint
extension for the operator $D_\a$: on its original domain of definition  
which consists of sections vanishing at both ends of the interval $[0,1]$, 
$iD_\a$ is symmetric only, while on sections satisfying periodic boundary 
conditions it becomes self--adjoint.    
\medskip 

The operator $D_\a$ is not invertible in the space of periodic sections.  
Its kernel is the ``diagonal part'' of the constant section. By ``diagonal''
we mean the subspace of the Lie algebra that commutes with $\a$, thus
being in the kernel of $\ad_\a$.  We therefore integrate over
$\phi^{diag}_0$ first. This produces a delta function $\d(J_0^{diag}) \equiv
\d\(\l^{diag}+(d\a)^{diag}\)$, which fixes the diagonal part of $\l$.  On the
remaining space the operator is invertible and we can perform the
fermionic Gaussian integration, using

\be \int \CD \psi  \exp \, \( \2 \psi_i \CO_{ij} \psi_j + \psi_i J_i 
\) =\Pf(\CO) \;  \exp \, \( \2 J_i \CO^{-1}_{ij} J_j \) \label{Gauss} 
\,\, . \ee
Here, $\psi$ and $J$ have been taken fermionic, the operator $\CO$ was
assumed to satisfy $\psi \CO \psi = - (\CO \psi) \psi$, and $Pf(\CO)$
denotes the Pfaffian of $\CO$. We get
\be 
I \ = \ \Pf(D_\a) \; \int d \widetilde \l \, \exp \, {\rm tr} \(\2 \sum_n 
J_{-n} D_n^{-1} J_n \right) \,\,,  \label{I2} 
\ee
where the Pfaffian is taken over the space of periodic sections without 
kernel and $\widetilde \lambda$ denotes the nondiagonal part of $\l$. 
Note that $J_0^{diag}\equiv 0$ so that the expression in the exponent is 
well--defined. Actually, since $D_n$ becomes merely the number 
$2\pi i \, n$ on diagonal elements, all of the diagonal parts of $J_n$ 
drop out due to $J_{-n}=J_n$ and the fermionic character of $J_n$. We 
indicate this again by means of tildes. Inserting the definition 
of $J_n$, eq.\ \re{I2}) becomes
$$
 I=\Pf(D_\a) \; \int d \widetilde \l \, \exp \, {\rm tr} \left[\2 \tl \(
\sum_n D_n^{-1} \) \tl + \tl (\ad_\a)^{-1} \wt{d  \a} + \2   \wt{d \a}
(\ad_\a)^{-1} \wt{ d  \a } \right] . 
$$
This is again a Gaussian integral for the variable $\tl$ and we assume 
that $\a$ is sufficiently ``generic'' for $\sum_n D_n^{-1}$ to possess
an inverse. We may again apply eq.\ \re{Gauss}) to obtain
$$ I =\Pf(D_\a) \,\Pf(\sum_n D_n^{-1}) \;  \exp \, {\rm tr} 
\bigg [- \2 \wt{d \a} 
\frac{ \displaystyle \(\sum_n D_n^{-1} \)^{-1}}{\displaystyle
\ad_\a^2} \wt{d \a} + \2 \wt{d \a} (\ad_\a)^{-1} \wt{d \a} \bigg ] 
\ , 
$$
where use of the  ad--invariance of the trace (Killing metric) has been 
made. 
\medskip

This result for $I = \int \CD \exp(\varpi_1)$ may now be combined 
with the expression for $\varpi_2$ in eq.\ \re{vp2}) to yield
$$ \int \CD \phi \exp(\varpi) =\Pf(D_\a) \,\Pf(\sum_n D_n^{-1}) \;  
\exp \left[ -\2 {\rm tr} \wt{d \a}  \frac{ \(\sum_n D_n^{-1} \)^{-1} 
+ e^{ \ad_\a}}{\ad_\a^2}  \wt{d \a}  \right] \  . 
$$
Here we made use of the fact that the diagonal parts of $d \a$ drop 
out in \re{vp2}) and that ${\rm tr}(\wt{d\a} \, f ( \ad_\a) \, \wt{d\a})$ 
vanishes for any function $f$ with $f(x)=f(-x)$. We are left 
with the computation of the operator $\sum_n D_n^{-1}$ and the two Pfaffians. 

We start with $Pf(D_\a)$. Denote by $i \a_r$ the nonvanishing
eigenvalues of $\ad_\a$, which are purely imaginary as $G$ is taken 
compact. The index $r$ runs over all roots in the Lie algebra of 
$G$; with $r>0$ ($r<0$) labeling the positive (negative) roots, 
one has $\a_r = - \a_{-r}$, furthermore.  In this notation one finds 
the following formal expression for the Pfaffian, 
$$
Pf(D_\a) = \( 
\prod_{r>0} i \a_r \) 
\prod_{n > 0} \( (2 \pi i \, n)^{\mbox{rank{\it G}}} \prod_{r'} (2
\pi i \, n + i \a_{r'}) \) \ \ .
$$
Clearly this is not well--defined. 
However, integrating $\exp \varpi$ over all of $\O G$ we cannot 
expect to obtain a finite result as the volume of the ``gauge group'' 
 $\O G$ is infinite. So we should divide (again formally) by this
 volume. The group of (based) loops is a group of even cohomology, 
$\e = \int_{0}^{1} ds \, {\rm tr} \( h^{-1}d h \, \partial_s  h^{-1}d h \) $ 
being the generator of $H^2(\O G)$. So, formally the Haar measure of $\O G$ 
is given by the infinite product of $\e$'s multiplied by the Haar measure 
on $G$ (since the zero mode drops out from $\e$). Using our previous notation 
and Fourier decomposition, $\e$ may be rewritten as ${\rm tr} \sum_{n \neq 0}  
(2 \pi i \, n) \, d \phi_{-n} d \phi_n$. Thus we are led to {\em define}\/: 
$$
\Pf(D_\a)/ \mbox{Vol} \O G := \( \prod_{r>0}  \a_r \) 
\prod_{n > 0}  \prod_{r'} \( 1 + \frac{\a_{r'}}{2 \pi \, n} \)
\,\,. 
$$
By means of $sin x = x \prod_{n=1}^\infty \( 1 - \frac{x^2}{n^2 \pi^2}
\)$ we then obtain
\be
\Pf (D_\a)/ \mbox{Vol}\ \O G  = \prod_{r>0} \( \sin(\frac{\a_r}{2}) 
\) \,\, . \label{Pf2} 
\ee
We remark that the square of this result agrees with the expression
obtained for $\det D_\a$ obtained in \cite{Cern} by means of zeta
function regularization.

We now come to the operator $\sum_n D_n^{-1}$, $D_n \equiv 2 \pi i \, n
- \ad_\a$, acting in that part of the Lie algebra that does not
commute with $\a$. Here we may use the simple formula 
$$    \sum_n \frac{1}{2 \pi i n - x} \ = \ \2 \coth(\frac{x}{2}) $$
to conclude that  
$$ \sum_n D_n^{-1} \ = \ \2  \coth (\ad_\a/2)  \ \ \mbox{ and thus } 
    \(\sum_n D_n^{-1}\)^{-1} \ = \ 2  \tanh (\ad_\a/2)\ \ .  $$
Putting all this together, we arrive at the following result:
$$ 
\frac{\displaystyle \int \CD \phi \exp(\varpi)}{\displaystyle 
\mbox{Vol}\ \O G } =  \prod_{r>0} \cos \( \frac{\displaystyle
\a_r}{\displaystyle 2} \) \exp \left[ -{\rm tr} \,   d \a \;
\frac{\displaystyle \sinh^3 (\ad_\a/2)}{\displaystyle
\ad^2_\a \cosh (\ad_\a/2)} \; d \a \right] \, \, . 
$$
Again, we have replaced $\wt{ d\a }$ by $d \a$ as the extra
contributions involving $ (d\a)^{diag}$ cancel anyway. We can 
finally rewrite the two--form in the exponent in terms of the
group element $\psi = \exp(\a)$. First, we remark that
$$
 \prod_{r>0} \cos \( \frac{\displaystyle
\a_r}{\displaystyle 2} \)=
\det^{\hskip 0.7cm \frac{1}{2}}\left(\frac{1+\Ad_\psi}{2}\right) \ \ ,
$$
where $\det^{\frac{1}{2}}$ denotes the unique 
positive square root of the matrix $(1+\Ad_\psi)/2$.
Next, the formula \re{lemm}) in Lemma 1
can be evaluated at $s=1$ to give $ d \psi  \psi^{-1}  = {\ad^{-1}_\a} 
(1-e^{\ad_\a}) d\a $. This may be inserted into our previous
result for the integral and leads to 
\be 
I(\psi)=\frac{\displaystyle \int \CD \phi \exp(\varpi)}{\displaystyle 
\mbox{Vol}\ \O G } =  \det^{\hskip 0.7cm \frac{1}{2}}\left(\frac{1+\Ad_\psi}{2}\right)
\exp \frac{1}{4}\left[ {\rm tr} \,  
d\psi \psi^{-1}\;
\frac{\displaystyle \Ad_\psi -1 }{\displaystyle
\Ad_\psi+1 } \; d \psi \psi^{-1}  \right] \, \, . \label{resulty} 
\ee

\section{Results and Discussion}

Combining equations (\ref{fiber}) and (\ref{resulty}), we obtain
the expression
\be
l\ =\ \det^{\hskip 0.7cm \frac{1}{2}}\left(\frac{1+\Ad_\psi}{2}\right)
 \exp \left(
\omega + \frac{1}{4} {\rm tr} \,  d\psi \psi^{-1}\;
\frac{\displaystyle \Ad_\psi -1 }{\displaystyle
\Ad_\psi+1 } \; d \psi \psi^{-1}  \right) \ \ .
\label{l}
\ee for the Liouville form $l$ on the orbit space $M$. 
The same expression was previously used in \cite{AMW} 
(formula (21)). Our path integral consideration gives a natural 
derivation of equation (\ref{l}), and shows its relation to the 
Liouville form $L$ on the field space $N$. 
\smallskip

Let us recall on this occasion that there was some freedom in our
computation associated with the choice of an anti--self--adjoint
extension for $D_\a$.  Instead of the periodic boundary conditions we
introduced in the paragraph below eq.\ \re{vp2}), we could have
extended the antisymmetric operator $D_\a$ also to sections with
different (only quasi--periodic) behaviour at the boundary. The 
final formula for $I(\psi)$ does depend on this choice of boundary 
conditions. It is expected, however, that the top degree part of 
the Liouville form $l$ in insensitive to this freedom in the 
computation. 
\medskip

Recall that the space $M$ arises as a result of reduction
from the field space $N$ with respect to  second--class
constraints. The residual first--class constraints $J^a_0$ generate
vector fields $v_a$ on $N$ which descend to $M$.  
According to \cite{AMW}, Proposition 4.1, the Liouville form $l$
satisfies the following interesting equation,
\be
\left( 
d + \frac{1}{24} f_{abc} \iota(v_a) \iota(v_b) \iota(v_c)
\right) l\ =\ 0\ \ .
\label{dl} 
\ee
Note that in the finite dimensional case of Section 2, 
$l=\exp(\omega)$ is a closed form. In the infinite dimensional 
situation we obtain an extra term $\frac{1}{24}
f_{abc} \iota(v_a) \iota(v_b) \iota(v_c)$ on the left hand side
of equation (\ref{dl}), which modifies the exterior differential
and should be interpreted as yet another manifestation of
the chiral anomaly. It is a very interesting
open question to trace the nature of this anomaly back to 
properties of the path integral in Section 4. 
\medskip

Simple examples of spaces $M$ are given by D--branes in the
WZW model \cite{AS}. There, the reduced spaces are conjugacy
classes in a group manifold, and the form $\omega$ is given by
the formula (see equation (7) in \cite{AS}),
$$
\omega \ = \ - \frac{1}{4} {\rm tr} \, \left(  d\psi \psi^{-1}\;
\frac{\displaystyle \Ad_\psi +1 }{\displaystyle
\Ad_\psi-1 } \; d \psi \psi^{-1} \right) \ \ .
$$
Formula (\ref{l}) shows that the form $\omega$ should
be corrected by the extra term arising from the 
path integral to yield 
$$
\tilde{\omega}\ =\ - {\rm tr}\, \left( d\psi \psi^{-1}\;
 \frac{1}{\Ad_\psi - \Ad_\psi^{-1}}  \; d \psi \psi^{-1} \right)\ \ .
$$
Note that in this case, the linear map $4^{-1}(\Ad_\psi -1)
(\Ad_\psi+1)^{-1}$ representing the correction term in eq.\ \re{l}) 
is inverse to the element $B = 4^{-1}(\Ad_\psi +1)(\Ad_\psi-1)^{-1}$ 
that appears in $\omega$. Hence, their difference $\tilde{\omega}$ 
is represented by $B-B^{-1}$. Surprisingly, the same combination 
shows up in the expression for the effective $B$--field derived
in \cite{S} in the analysis of D--branes on the flat background.
It is another challenging question to understand why the 
formula of \cite{S} applies to group manifolds and to 
establish the relation with the path integral of Section 3.
\smallskip

In this paper we did not touch the issue of quantization
of the spaces $N$ and $M$. While one can attempt to quantize
$N$ using the symplectic form $\Omega$, it is not clear what
it means to quantize $M$ because the form $\omega$ is not
closed. In the case of the D--branes in the WZW model
one can use the link between string theory and noncommutative
geometry to obtain an answer to this question \cite{ARS}.
The general case, however, remains an open problem.

\section*{Acknowledgement}

We would like to thank J. Kalkkinen for collaboration at
an early stage of this project and H.\ Grosse and E.\ Langmann 
for discussions and for their interest in our work. The hospitality 
of the Institutionen f\"or Teoretisk Fysik at Uppsala University, the 
Erwin Schr\"odinger Institute in Vienna and the II. Institut f\"ur 
Theoretische Physik at Hamburg University are gratefully acknowledged. 
This project was supported in part by a DAAD exchange program.

\end{document}